# Performance Modeling of WSN with Bursty Delivery Mode

Adel F. Agamy[a], Ahmed M. Mohamed[a]

[a]*Aswan University, Faculty of Engineering, Electrical Department,81542*

## Abstract

Wireless Sensor Network (WSN) usually consist of hundreds or thousands of sensor nodes scattered in a geographical area and one or multiple sink(s) collecting information. The special design and character of sensors and their applications make WSNs different from traditional networks. These characteristics pose great challenges for architecture and protocol design, performance modeling, and implementation. Accurately modeling the data generated by each sensor node is essential for correctly simulating network traffic, network congestion, interference between nodes and the energy expended by each node. Successful design leads to enhancing the overall performance of the whole of network. In this paper we analyze the performance of WSN with N-BURST traffic model. The impact of bursty traffic on the mean packet delay and buffer overflow probability is investigated analytically. We study the effects of bursty WSN traffic through simulations with three different cases. Both short-range dependency (SRD) traffic and long range dependency (LRD) traffic are simulated over different burst parameters. Finally we study the effect of pareto OFF time through simulation. The results are collected for 10 different 24 hour simulated periods in order to study and measure day-today statistical fluctuation.

## 1. Introduction

Wireless Sensor Network, as one of the most important techniques in the 21st century [1], widely used in facility safety, environmental monitoring and industrial applications. The WSNs are a new generation of telecommunication networks, which combine together the ability for sensing the environment with the possibility for local data processing and transmission through the wireless medium. These networks have many capabilities that make them suitable for a large variety of applications. Accurately modeling the data generated by each WSN node is essential for correctly simulating network traffic, network congestion, interference between nodes and the energy expended by each node. Understanding traffic flow in a WSN lead to successful design of WSN. In the WSN literature, the performance evaluation of the protocols are generally carried out with periodic data traffic as in [2], or using common data traffic models such as Poisson point processes [1, 3, 4, 5]. However, event-driven applications such as target detection and tracking produce bursty traffic which cannot be modeled as either periodic or Poisson [6, 7, 8]. Although there are packet traffic models available for legacy communication networks, the unique features and requirements of WSNs call for the design and development of dedicated models. Constructing accurate and analytically tractable source models for sensor network traffic will provide a basis for further work on proposed network protocols. Performance evaluation of WSNs will be performed with realistic traffic loads. Besides, the effects of system parameters can be analyzed without the need for simulations. The main contributions of this paper can be summarized as follow

- To the best of our knowledge, we are among the first to present analytical tractable model for WSN and verfied it by simulation.
- Present a general traffic modeling for WSNs.
- Model the traffic of each node by 1-burst model.
- Provide a general formula for mean packet delay and buffer overflow probability under different application.
- Study the effect of different application on the performance of WSN.

The rest of the paper is organized as follow. Section 2 mentions the related work. Section 3 define N-Burst traffic model and show how it is more general to model most types of traffic. In section 4 we define WSN model by N-Burst and define the parameter of single node. Section 5 shows by analytical analysis the behavior mean packet delay and buffer overflow probability with different burst parameter. Section 6 shows the simulation result with three different case study. In section 7 we conclude our work.

## 2. Related Work

Several approaches have been proposed for the characterization of the performance of WSNs. Deterministic worst-case performance analysis is introduced in [9] using deterministic traffic model for real time application. In [10], the authors

*Email addresses:* `a.f.agamy@aswu.edu.eg` (Adel F. Agamy), `ahmed@engr.uconn.edu` (Ahmed M. Mohamed)

propose analytical model for calculating RRT (Reliable Real-Time) degree in multihop WSN and they calculated a worst case analysis of the delay and queue quantities in sensor networks using network calculus. In [1] authors present approach to analyze data traffic flow in the mac layer of WSN with exponential distributed service time ans with Poisson distribution packet arrival. In [8] bursty traffic was generated for certain time of interval to adaptive MAC protocol for WSN. In [11], an analytical traffic flow model is developed for cluster-based WSN. The source-to-sink path is modeled by a number of single-server finite queues linked in tandem and they modeled each node by a queue with poisson arrival process and phase distribution for service time with finite buffer. In [12], stochastic data traffic model is introduced for medical WSN; the authors modeled the traffic generated by a single WSN node monitoring body temperature and electrocardiogram (ECG) data.T hey found that the traffic profile values are distributed according to X, where . indicates magnitude and X is a zero mean Gaussian random variable with a variance that changes with time. They compare the autocorrelation of a compressed MIT_BIH waveform to that of a traffic profile generated with their model. In [13], sensor network packet traffic model is derived and analyzed for intrusion detection applications. In [14], the performance of bulk data dissemination in Wireless Sensor Networks is analyzed. The traffic traces of an intrusion detection scenario are studied in [15] where numerical function fitting is carried out for the total number of packets generated at any instance. However, no generalized analytical packet traffic model is derived. An analytical model for intrusion detecting WSN is investigated in [6]. In [16] the authors propose a Markov chain-based analytical framework for modeling the behavior of the medium access control (MAC) protocol in IEEE 802.15.4 wireless networks. They present two scenarios. First, they consider networks where the (sensor) nodes communicate directly to the network coordinator (the final sink), in this scenario the nodes generate traffic with the same distribution, GEO/G/1/L queue model is used to represent the network. Then, they consider cluster-tree (CT) scenarios where the sources communicate to the coordinator through a series of intermediate relay, which forward the received packets and do not generate traffic on their own). In this scenario the network can be modeled as a two coupled GEO/G/1/L queue model, using DTMC (Discrete Time Markov Chain) model to determine the throughput and delay. In [17], the authors show the significance of using a realistic and application-specific packet traffic model by comparing the performance of a well-known WSN protocol under the Surveillance WSN packet traffic model (SPTM), as well as under periodic and binomial traffic models. The authors in [18, 19] suggest that the packet arriving sequence at the place of individual sensor nodes can be modeled, and this can be conducted by extracting all given length unique subsequences observed during a sufficiently long time span and constructing a pattern database using these extracted subsequences. Following on from this, the pattern database can be used as the traffic profile for the corresponding node. For an event-driven sensor network, bursty traffic can be triggered upon detection of an interesting event. Traditional Poisson processes have been shown to be inappropriate in modeling the bursty traffic [12, 13]. The author proposes in [19, 20] the use of an ON/OFF model to capture the burstiness of communication traffic in event-driven WSNs. Furthermore, the ON/OFF period distributions and their properties are studied. In [20], source traffic dynamics in a simulated target tracking WSN scenario are explored. The author found that the source traffic arrival process does not follow the usually considered Poisson model. Instead, an ON/OFF model is found to be capable of capturing the burst nature of the source traffic arrival. In addition, they found that the measured ON/OFF periods follow the generalized pareto distribution. Mathematical analysis also shows a surprising fact: all ON/OFF period distributions in the experiment exhibit a short-tail property, which is a nice property that could be exploited by applications such as anomaly detection and node failure detection. But the authors don't study the behavior of mean packet delay or buffer over flow probability. The author in [21] use two data generation model to show the effect of data aggregation on the performance of WSN, the first each node generate a traffic that follows a Poisson process with rate $\lambda$ ,the second each node generate traffic follows an ON/OFF bursty process where packets are only generated while the process is in the ON state. For the ON/OFF the authors consider two different traffic modes. For traffic mode 1, the duration for which the process stays in the ON and OFF states follows exponential distribution with mean 2 and 8, respectively, while for traffic mode 2, the duration for ON state is uniformly distributed between 1 and 50 and the duration for OFF state is exponentially distributed with mean 50 seconds. None of the previous models capture the heavy tail traffic, self-similar, bursty traffic. Our model use N-BURST traffic model to model the traffic generated by WSN nodes. Performance metrics such as mean packet delay and buffer overflow probability is obtained analytically and verified by numerical result under different burst parameter (traffic shaping) and utilization. A truncated power tail distribution (TPT) is used to model the ON time distribution, with reliability function as follow.

$$R(x) := \frac{1-\theta}{1-\theta^T} \sum_{j=0}^{T-1} \theta^j \exp\frac{-\mu x}{\lambda^j} \quad (1)$$

T = 1 corresponds to the exponential distribution. Where it used very successfully in standard queueing models[22, 23, 24, 25, 26]. For full details, see [27] where the effect of different truncation parameter is investigated.

## 3. N-BURST TRAFFIC MODEL

In [28], the authors found that the N-BURST traffic model can capture the traffic characteristics of most types of computer networks. The N-Burst introduced in [22, 23] is a variant of the many ON/OFF models described in literature. The N-Burst arrival processes is a superposition of traffic streams from N independent, identical sources of ON/OFF type. During its ON-time each source generates packets at rate $\lambda$ b and it is quiet during its OFF time. This arrival process, with arbitrary ON and OFF time distributions (having Matrix-Exponential (ME) representations, see 26) is analytically modeled as a Semi-Markov process of the Markov Modulated (MMPP) type. The details of this Poisson Process model can be found in[22, 23]. When using Power-Tail distributions for the duration of the ON periods, self-similar properties, which are critical for understanding bursty traffic is generated. Accommodating both burstiness and self-similarity in an analytic point-process model is not easy and thus many approximations have been used by various researchers to understand buffer overflow problems and packet delay. Some examples include the M/G/1 queue where the service time has infinite variance [29], continuous flow models during bursts[30, 31], and batch arrivals. The burst models are also known as ON-OFF models. For very low intra burst packet rates, the N-Burst/G/l model reduces to an M/G/l queue. For $\lambda b \to \infty$ all packets in a burst arrive simultaneously and the model becomes a Bulk arrival.

## 4. WSN network model Terminology

The applications and corresponding traffic characteristics in WSNs are different from those of traditional networks. As a result, traffic and data delivery models are also different. Currently, four traffic models are used in WSNs: event-based delivery, continuous delivery, query-based delivery, and hybrid delivery. Traffic modeling greatly influences protocol design and affects performance. In continuous delivery the data collected by the sensors need to be reported regularly, perhaps continuously, or periodically and is modeled in [2]. In query based delivery sometimes, the sink may be interested in a specific piece of information that has already been collected in sensor nodes, it asks nodes for this information then nodes send information. In event-based delivery sensor nodes monitor the occurrence of events passively and continuously, in this case the traffic depends on the distribution of the phenomenon. WSN model consists of N sensor nodes, each generates data according to 1-Burst traffic model and only one powerful sink node that collects data generated by all of sensor nodes. We assume that WSN nodes are static and nodes can directly communicate with the sink node as illustrated in fig 1. We also assume exponential service time and infinite buffer sink node. Now we define the single node traffic model using N-Burst traffic model:
K: = the mean arrival rate for each node (the average for ON- and OFF-times together) $\lambda$ := Overall arrival rate (packets per time unit) that generated by N-node Where $\lambda$ =KN.
$n_p$:= Mean number of packets during a burst(active period).
$\lambda p$:= peak transmission rate during a burst (packets per time unit).
ON:= $n_p / \lambda_p$ = Mean ON (active period) time for a burst (time units).
OFF:= Mean OFF(sleep period) time between bursts (time units).
v:= Mean packet service rate of sink node (packets per time unit).
$\rho$:= $\lambda$/v = sink node utilization.

In addition, the burstiness parameter, b, defined as:

$$b = \frac{\overline{OFF}}{\overline{ON} + \overline{OFF}} = 1 - \frac{K}{\lambda_p} \qquad (2)$$

This parameter can be thought of as a shape parameter. Our traffic model depends on four separate distributions as introduced in [30, 22], Each one governs a different sub-process which all together describe characteristics of the 1-Flow model. The important ones are: OFF: OFF-Time distribution with mean OFF (depends on how bursts are generated, and how often) ON: ON-Time distribution with mean ON causing a mean number of $n_p = \lambda_p$ ON a packets in a burst.

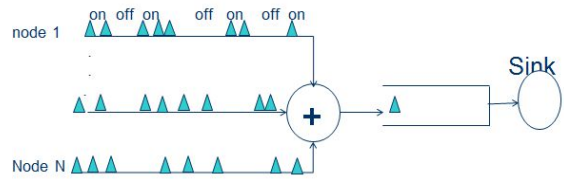

Figure 1: WSN Model

## 5. ANALYTICAL ANALYSIS

### 5.1. performance metrics

Although the analytic N-node WSN traffic network model can be evaluated for all possible distributions (within the ME class), the calculations are not trivial. It is possible to get much insight by looking at the limiting cases for b = 0 and b = 1. It is not hard to see then, that for mean packet delay (mPD) (as well as for buffer overflow probabilities) b = 0 corresponds to the best performance (smallest mPD), while b = 1 corresponds to worst case (largest mPD). Furthermore, the mPD is a monotonically increasing function of b. Therefore, it is useful to discuss some well-known formulas in queuing theory. For b = 0 the SM /M /l queue reduces to the $M_\lambda/M_v/l$ queue, in which case, the mean packet delay is given by the elementary formula

$$mPD(b=0) = (\frac{1/v}{1-\rho}) where \rho = \lambda/v \qquad (3)$$

At the other extreme (b = 1) is the bulk arrival queue. This behavior is also well know (see, [27]), and can be written as:

$$mPD(b=1) = (D\frac{1/v}{1-\rho}) where D = \frac{E(\frac{L(L+1)}{2})}{E(L)} \qquad (4)$$

Where L is random variable counts the number of packets in active period. In our parametric studies to be described below, we will hold $\rho$ constant, while varying b, for various ON-time distributions. That is, the average load is held constant, while the packets in a burst are bunched up or spread out as much as possible according to the value of b. in this paper we use the N-bursts/M/1 queue model with ME representation of N-bursts and calculate analytically the steady state mean packet delay and buffer overflow probability for N=1,2.

## 5.2. Results for a WSN with one node

We have chosen to look first at the WSN with one node, with fixed utilization parameter, $\rho$, and exponential distributions for all but the ON-times. The ON-times are distributed according truncated power tail distribution (TPT). This class of functions is robust enough to demonstrate the qualitative differences among distributions with different variances. In all cases presented here, we set $\theta = 0.5$, and $\alpha = 1.4$. T = 1 corresponds to exponential ON-times (or equivalently, geometric distribution of number of packets in a burst). fig 2 shows the mean packet delay as a function of b, the Burstiness parameter as given in equation (2), for various values of T. All points on this graph have the same value for $\rho = 0.5$. That is, in all cases the sink is busy only 1/2 the time. As can be seen in fig 2 when b is small enough, delay is negligible and is insensitive to the ON-time distribution. In this case $\lambda$p is small (close to $\lambda$), so the packets of a burst are approximately spread over the whole time between burst starts. This is not necessarily desirable, since there may be considerable delay between the first and last packets of a burst. As the packet rate is increased, thereby decreasing the time for the source to transmit a burst, the mPD increases gradually to the bulk-arrival limit at b = 1, but only if the ON-time distribution is well behaved (e.g., when T≤10). The behavior of mPD changes when truncated Power Tail distributions with larger truncation parameter T are considered. As shown in fig 2 when b increase, the mPD increase smoothly but at b=0.5 the mPD jumps by two or more orders of magnitude exceeds the service rate V. as can be seen in fig 3 the buffer over flow probability approximately zero before the b=0.5, at b=0.5 the buffer over flow probability jumps only with large T. In fact, as T→ ∞ the jump becomes unboundedly large. By the definition of b, the jump occur when $\lambda_p$ >v, the condition $\lambda_p$ > V corresponds to b > (1 −$\rho$). The point, $b_l$ = 1- $\rho$, is called the blow-up point. The region b > (1 −$\rho$) is called blow-up region 1. For N > 1(more than one source), several blow-up regions exist. It is clear then, that if the distribution of number of packets in a burst corresponds to a distribution with large T, then sink performance will be unacceptable in the blow-up region. This is true for all $\rho$. fig 6 illustrates, how the location of the blow-up point depends on the utilization, namely that blow-up occurs near $b_1$ = 1- $\rho$.

## 5.3. Results with two nodes model

When two or more node supply bursts to a sink, the mPD structure becomes more complicated particularly if the ON-time is TPT distributed with large T. We only supply two nodes analysis only. First consider the case b = 0. Then, as with the one node WSN, each node is a Poisson process if the intra−burst distribution, $X_{IN}$, is exponential. As is well known, the merging of several Poisson Processes is still a Poisson Process whose rate is the sum of the individual rates [32]. Thus, for small b, behavior is as simple as one could hope, for any number of sources result in an $M_\lambda/M/l$ model where $\lambda$ = NK. We mention here that if the intra-burst distributions are not exponential, then the collective process is not a renewal process. But if the distributions are well behaved, then mean packet delay should be reasonably low for small b [22], even if the ON-Time distribution is PT. At the other end, where b = 1, if the OFF-Time distribution is exponential, then each source yields a Poisson process of bulk-arrivals. Again, the merged process is also Poisson, yielding the same performance as the one WSN node limit, with cumulative bulk arrival rate. If the OFF-Time distribution is not exponential, but still well behaved, then we would expect performance comparable to the exponential case. The above discussion asserts that sink behavior in the regions near b = 0 and b = 1 is the same independent of the number of sources. This is shown in fig 2 and fig 4 where our results for a two WSN nodes calculation are presented. It is necessary to compare the one and two node WSN processes with some care. To maintain the same load on the sink node, the one node WSN model submits packets at twice the rate of each of the sources of the two node model ($\lambda$ = 2k) furthermore, in order to have the same. Burstiness parameter, the one WSN node must have a peak transmission rate that is also twice that of each of the two nodes WSN model. In general, for an N-node WSN model:

$$b = 1 - \frac{k}{\lambda_p} = 1 - \frac{\lambda}{n\lambda_p} \qquad (5)$$

fig 2 and fig 4 shows clearly that for small b, sink performance is largely independent of $X_{IN}, X_{OFF} and X_{ON}$. Similarly, for b close to 1, it is again independent of N and $X_{IN}$. But the structure of the blow-up regions is very dependent on N. We see that at the leftmost blow-up point ($b_2$ = 1-$\rho$ = 0.5) the two WSN nodes do not exhibit nearly so large a jump as the one node. Part of the reason is that both sources must simultaneously be in a long burst for the blow-up to occur. The other reason is that the distribution of this double event is power-tailed with exponent $2\alpha − 1$, which is larger than $\alpha$.

## 5.4. Blowup points locations

For large T, fig 4 shows a second blow-up point for the WSN with two node, at $b_1 \approx 2/3$. For two sources, Blow ups occur when the sum of the average of one source together with the other being always ON is sufficient to saturate the sink. That is, when v = $\lambda$ p + k translates to $b_1$ = 1-$\rho$ / (2 − $\rho$), which for $\rho$= 0.5 is $b_1$ = 2/3. In general, the location of the N blowup points is given by [33]:

$$b_i = N\frac{1-\rho}{N-\rho(N-i)} for i = 1,2,3...N \qquad (6)$$

The blow-up points are located by the condition: V= $A_i$ Where $A_i$ = i $\lambda_p$ + ( N − i )k, which leads to the above formula for $b_i$. We call the interval $b_i < b < b_{i-1}$ the blow-up region i .

## 6. simulation

### 6.1. Model Setting

To clearly understand system behavior of a complicated WSN model it was necessary to first begin our study on a simple WSN networks with N node and one sink where the node communicate directly with sink. Here packets generated by the N ON-OFF nodes arrive at a sink where they are queued in an infinite buffer until they can be sent to their destination (see fig 1).

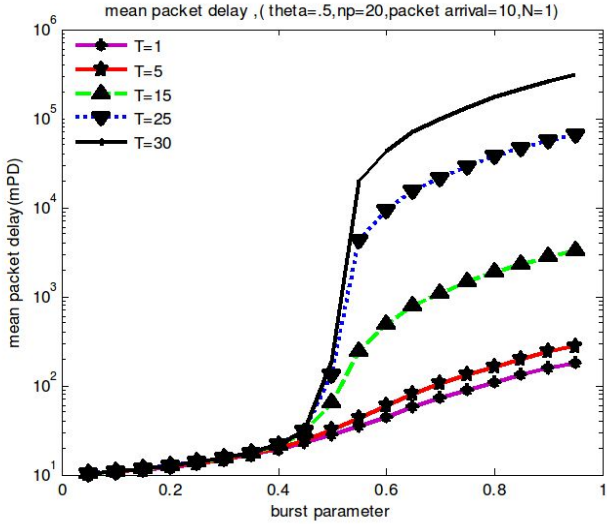

Figure 2: the mean packet delay(log scale) with different ON time distribution.

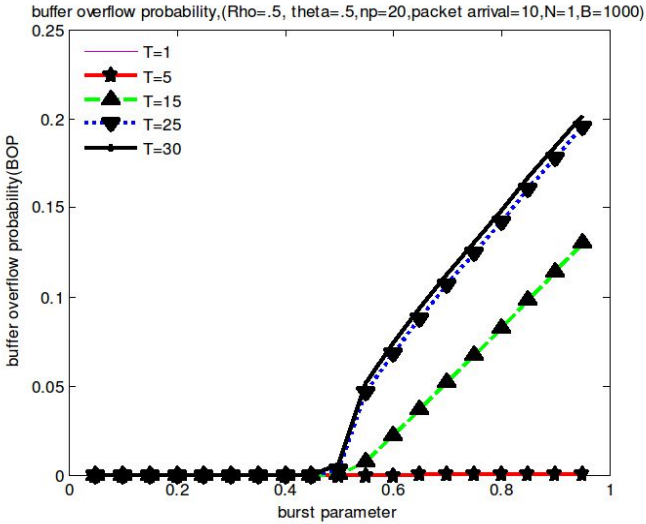

Figure 3: The buffer overflow probability with different ON time distribution

We assume that the packets are variable in size according to an Exponential distribution with mean of 100 byte. There are on average $n_p = 50$ packets transmitted with constant peak rate during a burst. Two distributions: Exponential (EXP) and pareto (PT) were chosen to represent the distribution of the number of packets in a burst. There are many versions of pareto distributions presented in the literature. Let X be the random variable denoting the number of packets during an ON state. Then

$$R(x) = \frac{1}{(1 + \frac{x}{M(\alpha-1)})^x} \qquad (7)$$

Where R(x) = Pr(X > x), is a reliability function, $\alpha$ is a shape parameter, whose value is 1.4 in this study, and M is the mean of the distribution, (i.e., E(X) = M). EXP produces short-range dependency (SRD) traffic, while PT yields long-range dependency (LRD) traffic. Time between bursts (OFF-time distribution) is exponentially distributed for the the two types of traffic. The third type of traffic we represent both OFF time and ON time distribution with PT. The sink has speed, v= 100 packets per second and transmits packets in a first-in first-out, (FIFO) manner. Our goal is to study the behavior of bursty traffic using the NFlow model and its impact on WSN networks performance such as mean packet delay and throughput of the network under constant sink utilization. Each nodes sends data according to some ON-OFF pattern for the entire simulation. The number of nodes at any time is the same. Since it is impossible to simulate a flow of infinite length, we truncate our simulations after a 25 hour period and collect relevant statistics after the warm up period of the first hour. Both SRD and LRD traffic patterns are simulated at different values of b and their impact on network performance is studied. We evaluate our study through simulation under a case study below.

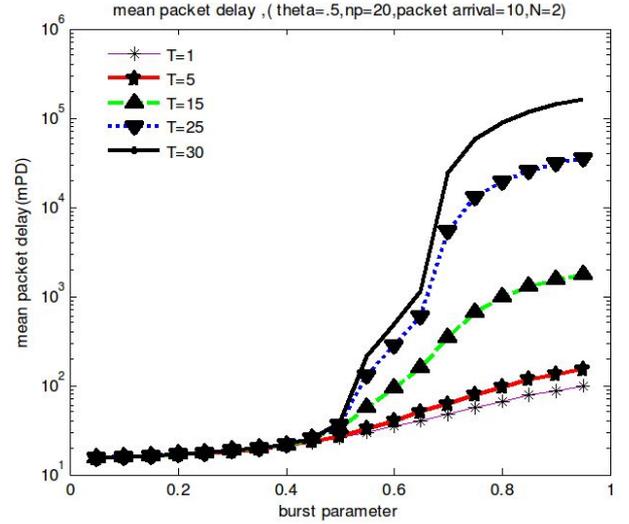

Figure 4: The mean packet delay of two node model with different ON time distribution

### 6.2. Case Study1

We designed this case study such that a sink sees the same utilization, $\rho$, regardless of number of active nodes, N. Hence, for any $N \geq 1$, the overall packet arrival rate, $\lambda$, as seen by the sink, is the same (chosen as 50 packets/second in this case)even though the mean packet rate of individual nodes, K, can be different depending on the value of N. Recall that k = $\lambda$/N. Thus, the more nodes being active, the smaller packet rate each nodes carries. The key idea here is to test if a sink could differentiate the differences in number of active nodes and how these changes affect the network performance. In addition, for a given N, we complicate the scenarios even more by varying $\lambda$p of each node while keeping its K the same. Therefore, even though the aggregated traffic seen by the sink consists of the same number of individual active nodes, each with the same K, the sink may experience the differences in the traffic shapes. This depends on $\lambda$p because changing the peak rate also changes the burst parameter (See Equation 2). The goal here is to study how the sink responds to such differences in traffic behavior and ultimately how it performs as the number of active nodes increase. We simulate traffic of the N ON-OFF nodes, where N =1, 2, 5 and 10. For LRD traffic, the ON-time distribution is defined by a pareto distribution with $\alpha = 1.4$. For SRD traffic, the ON-time is exponentially distributed with the same mean. Also we study the effect of the pareto OFF time distribution on the performance by simulating the model with N=1. For a given number of active nodes, burst parameter, b, varies from 0.05 to 0.95 for each run. For a given number of active nodes, burst parameter, b varies from 0.05 to 0.95 for each run. Performance metrics used in this study are described as follows: 1) Mean Packet Delays, the average time a packet takes to go from its node to a sink. 2) Throughout, the number of packets that arrive at the sink after the warm up period (within 24 hours), divided by the period of simulation (24 hours).

### 6.3. Simulation Results of case 1

The results presented in this section validate the analytical results shown above. fig 7 and fig 8 present the performance fluctuation of 10 different days for EXP traffic when N =1, 2,5 and 10. fig 9 and fig 10 present the performance fluctuation of 10 different days for pareto traffic when N=1. The results show that PT traffic causes much more fluctuation and worse performance than EXP traffic especially in term of delays. This indicates that power tail traffic such as pareto cause not only worth but also more unpredictable performance from day to day. However, the fluctuation reduces when N increases, as illustrated in fig 11 and fig 12 where N = 10. It was clear that PT traffic causes much worse performance than EXP traffic. When we change the OFF time to pareto the result show much more fluctuation and worse than pareto traffic with exponential OFF time as in fig 13 and fig 14. The results suggest that a network performs better when the total stream of packets is distributed over more nodes. Hence, a network yields the least average delay at N = 10, but with the throughout of the system is low. Recall that the higher number of active nodes that exist will lower the average rate of each node. Hence, overall traffic is smoothed out when N is higher which helps a network to provide better performance. The shape of the graphs also reflects the effect of the burst parameter on network performance. The higher the burst parameter is, the higher the peak rate of each node carries. Whenever the sum of all arrival rates exceeds the sink rate, performance of the network decreases. Therefore, a network performs worst when b is closer to 1. Note that for PT traffic graphs of one active node in fig 9 and fig 10 start blowing up at b = 0.5 and continue increasing as b increases. For N > 1, their several blow-up points which calculated from section 5.4. Let us consider the impact of different numbers of sources at the same blowup location. Note that at the leftmost blowup point (b=0.5), figures of the one active node show the highest jump followed by the 2nodes and so on. This is because to have a blowup at the leftmost location for any N >1 all N sources must be in an ON state simultaneously in order to meet a saturation condition as in section 5.4 and chance of this event to occur is smaller than of the one active node. In addition, we see that at other blowup locations beyond b=0.5, figures of the one active node yield the poorest performance followed by of the two and so on. This is because when b>0.5 the one active node already resides in its 1st blowup region while other N sources are still gradually moving toward their 1st blowup region as b increases. Note that N=10 is the last one arriving at its 1st blowup point at b≈0.9 as in fig 11 while N=1 have been in 1st regions for a while and their transmission rates already significantly exceed the sink rate ($\lambda_p >$v). That is why they have greater impact on the performance than N=10.

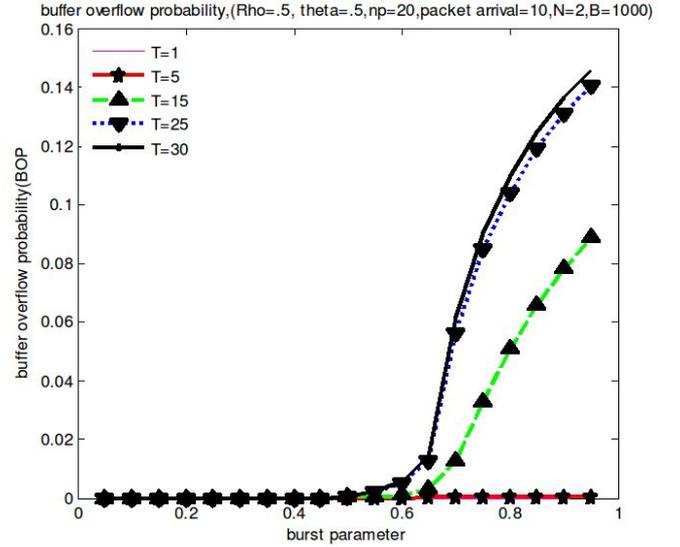

Figure 5: The buffer overflow probability of two node model with different ON time distribution

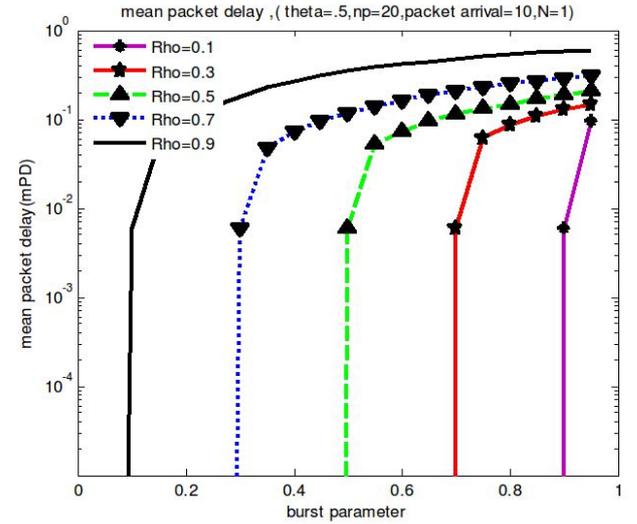

Figure 6: The effect of utilization($\rho$) on the location of blowup points

### 6.4. Case study 2

In this case we consider a general cluster tree network, where the traffic generated at the sources nodes flows towards the sink, through a series of intermediate relays. In particular, each relay receives packets coming from a specific cluster of sources. At the same time, the relays may be grouped into higher level clusters, which can be associated with even higher level relays or by the sink itself. By grouping the sources and the relays at the various hierarchical levels, one obtains the cluster tree network depth. Our case study assumes the network topology as in fig 15 is a particular cluster tree network with a depth equal to 3 and the following hierarchical levels: the first level contains two clusters each one has N sources, the second level is composed by the single relay for each cluster , and the last level contains just the sink. We designed this case study such that the sink and relays see the same utilization, $\rho$, regardless of number of active nodes in each cluster. The overall packet arrival rate, $\lambda$, as seen by each relay, is the same (chosen as 50 packets/second in this case) even though the mean packet rate of individual nodes, K, can be different depending on the value of N in each cluster. Recall that k = $\lambda$/N, thus, the more nodes being active, the smaller packet rate each nodes carries. . We simulate traffic of the N ON-OFF nodes, where N =1, 2, 5, and 10. For LRD traffic, the ON-time distribution is defined by a pareto distribution with $\alpha$ = 1.4. For SRD traffic, the ON-time is exponentially distributed with the same mean. For a given number of active nodes, burst parameter, b, varies from 0.5 to 0.95 for each run. Performance metrics used in this study are described as follows: 1) End to End Mean Packet Delays, the average time a packet takes to go from its node to a sink (for cluster 1and 2), 2) Throughout of relay 1 , the number of packets that arrive at the relay 1 from cluster 1 after the warm up period (within 24 hours), divided by the period of simulation (24 hours), 3) Throughout relay 2, the number of packets that arrive at the relay 2 from cluster 2 after the warm up period (within 24 hours), divided by the period of simulation (24 hours). 4) Throughout of the sink, the number of packets that arrive at the sink after the warm up period (within 24 hours), divided by the period of simulation (24 hours).

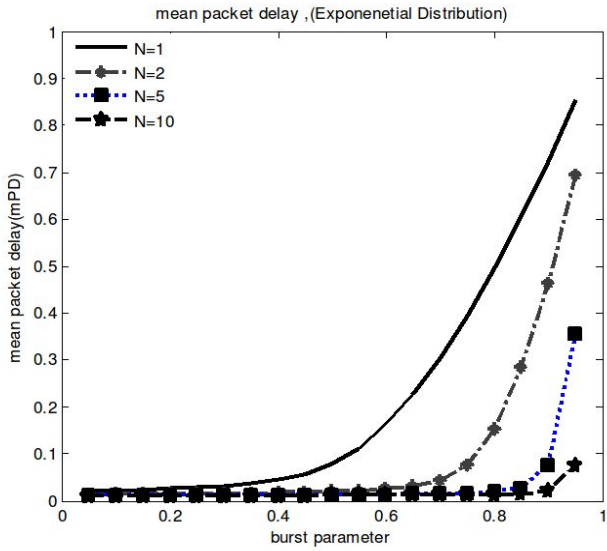

Figure 7: The mean packet delay with EXP ON time distribution at different N

*6.5. Simulation Results of case 2*

The results presented in this section show that the behavior of end to end mean packet delay of cluster 1 and cluster 2 is the same as in the case study 1 as in fig 16 for exponential traffic and in fig 18 and fig 20 for LRD traffic. This means that the relays have not much effect on the end to end delay of cluster 1and cluster 2, and the delay is dominated by the burst parameter and the number of active nodes in each cluster. The throughout of the sink is approximately the sum of the throughout of cluster1 and cluster 2 as in fig 17 for exponential traffic and in fig 19 and fig 21 for LRD traffic. This means all packets departed from relays 1and 2 reach the sink. The overall result show that the relays have not affected on the end to end behavior of bursty traffic. This because as the number of relays increase, the arrival traffic at each relay became smooth (approach Poisson arrival).

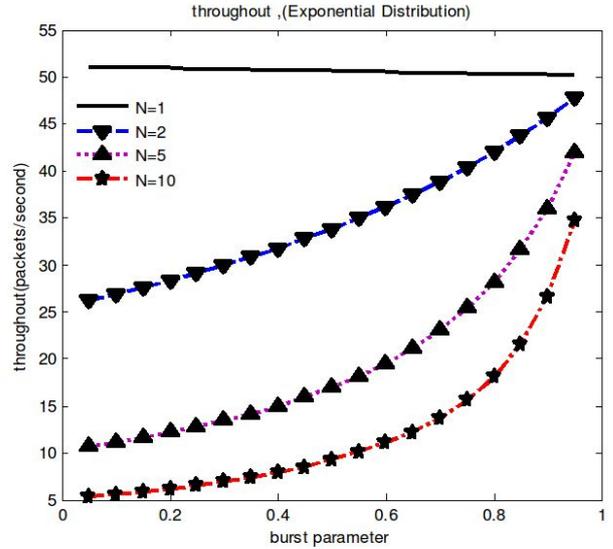

Figure 8: The throughout with EXP ON time distribution at different N

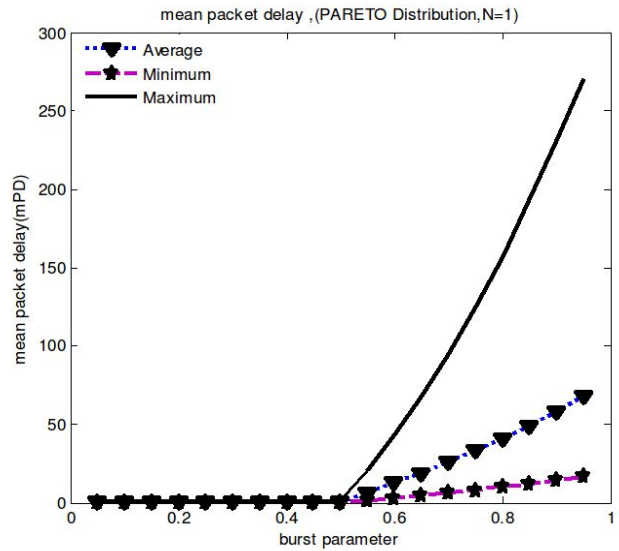

Figure 9: The mean packet delay with PARETO ON time distribution with N=1

*6.6. Case study 3*

Our case study assume the network topology as in Fig fig 22 is a particular cluster tree network with a depth equal to 3 and the following hierarchical levels: the first level contains two cluster each one have N sources, the second level is composed by the single relay for each cluster, and the last level contains one cluster and the sink, where the node in the cluster can communicate direct with sink. We designed this case study such that the sink and relays see the same utilization, $\rho$, regardless of number of active nodes in each cluster. The overall packet arrival rate, $\lambda$, as seen by each relay, is the same (chosen as 50 packets/second in this case) even though the mean packet rate of individual nodes, K, can be different depending on the value of N in each cluster. Recall that $k = \lambda/N$. Thus, the more nodes being active, the smaller packet rate each nodes carries. We simulate traffic of the N ON-OFF nodes, where N =1. For LRD traffic, the ON-time distribution is defined by a pareto distribution with $\alpha = 1.4$. For SRD traffic, the ON-time is exponentially distributed with the same mean. For a given number of active nodes, burst parameter, b varies from 0.5 to 0.95 for each run. Performance metrics used is the same as the previous section 6.5. The aim of this case to study the effect of the cluster 3 which communicate direct to sink on the end to end delay of cluster 1and 2.

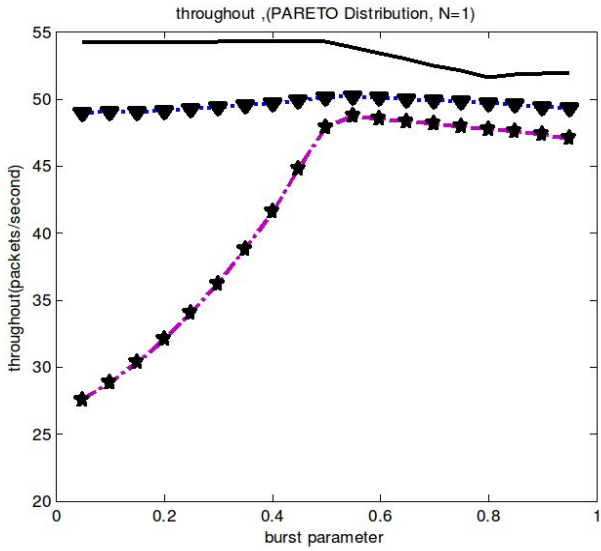

Figure 10: The throughout with PARETO ON time distribution with N=1

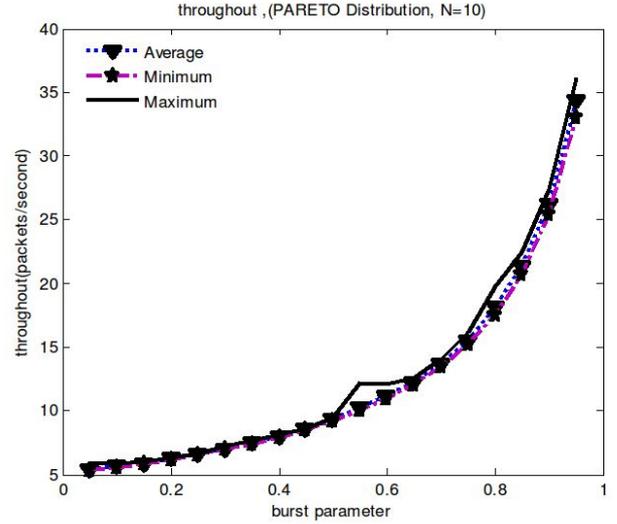

Figure 12: The throughput with PARETO ON time distribution with N=10

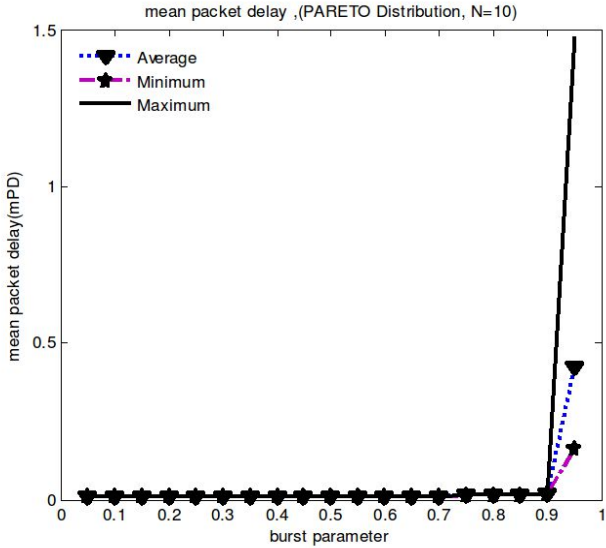

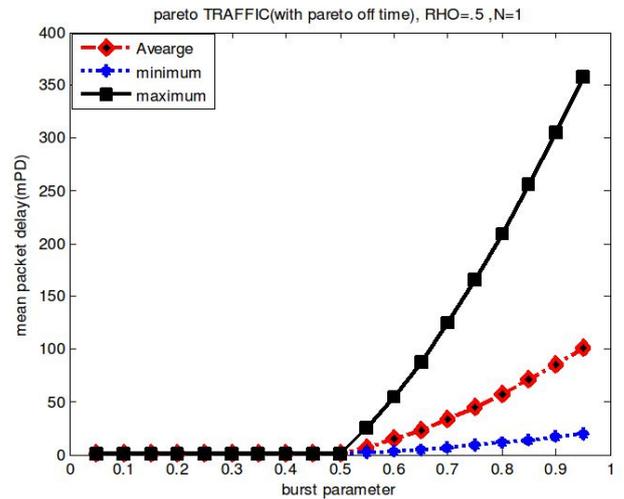

Figure 13: the mean packet delay with PARETO ON time and OFF distribution with N=1

Figure 11: the mean packet delay with PARETO ON time distribution with N=10

### 6.7. Simulation Results of case 3

The results of presented in this section show that the behavior of end to end mean packet delay of cluster 1 and cluster 2 is the same as in the previous case study (without cluster 3) as in fig 23 for exponential traffic and in fig 25 LRD traffic but with larger delay. This means that the existence of the third cluster have not much effect on the end to end delay of cluster 1 and cluster 2 when b small, the effect jump when the arrival rate of the third cluster exceed the service rate of the sink. The throughout of the sink is larger the sum of the throughout of cluster1 and cluster 2 as in fig 24 for exponential traffic and in fig 26 for LRD traffic. The overall result show that the existence of the third cluster have affected on the end to end behavior of bursty traffic only when the third cluster enter the first blowup region (the arrival rate of the third cluster exceed the service rate of the sink).

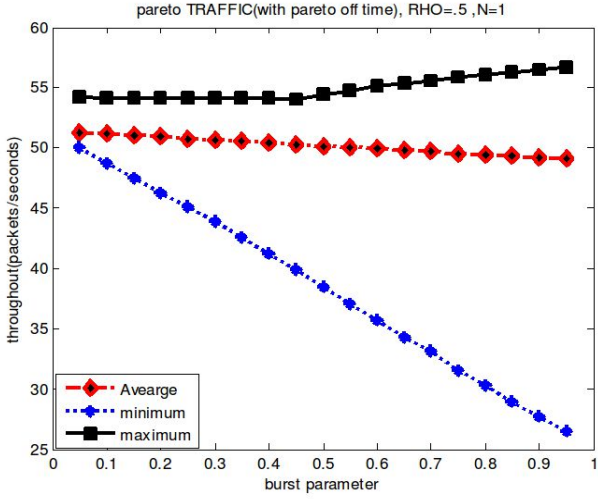

Figure 14: the mean Throughput with PARETO ON time and OFF distribution with N=1

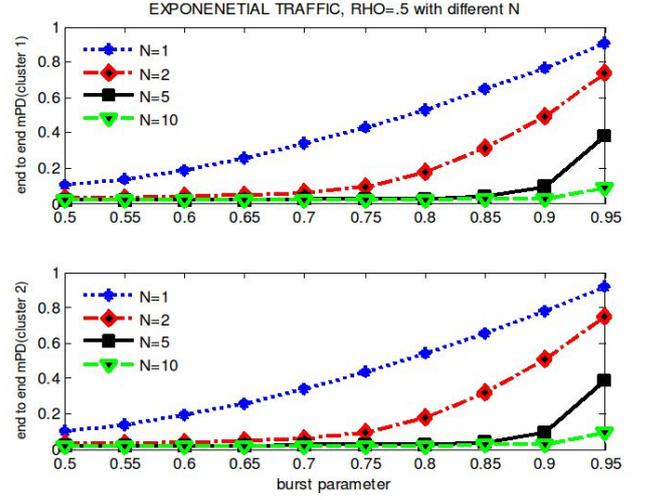

Figure 16: The mean packet delay with EXP ON time distribution with different N

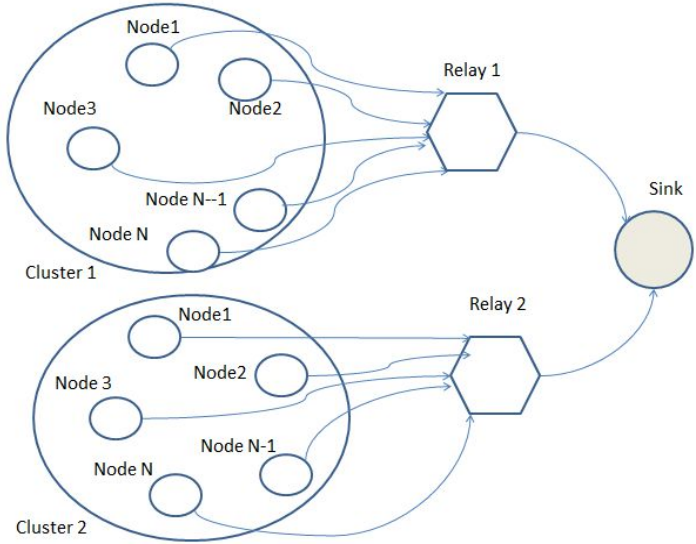

Figure 15: Cluster Wireless Sensor Network

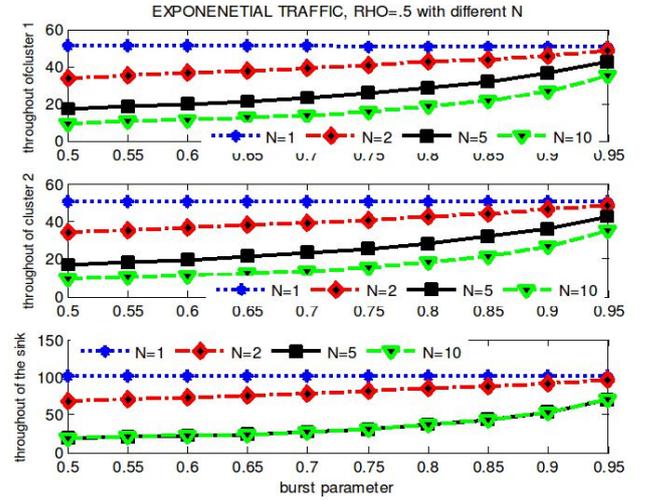

Figure 17: The throughout with EXP ON time distribution with different N

## 7. conclusion

The WSNs are a new generation of telecommunication networks, which combine together the ability for sensing the environment with the possibility for local data processing and transmission through the wireless medium. These networks have many capabilities that make them suitable for a large variety of applications. Understanding traffic flow in a WSN lead to successful design of WSN. We use N-Burst traffic model for representing wireless sensor network traffic fed to sink node. By introducing Power-Tail distributions for the distributions of ON-times the N-Burst model is able to represent those traffic characteristics that dominantly determine performance of WSN components. The reason for using TPT distributions in our model is twofold: (1) TPT distributions have a matrix exponential representation and thus allow matrix geometric/algebraic solutions; and (2) make it possible to do parametric evaluation not only w.r.t. the power index a but also to vary the T over a range of interest corresponding to the fitting to real data. An example of such a parametric evaluation is displayed in fig 2, it shows the dramatic increase of mean packet delay in the region of high burstiness. Setting the parameter b reflects the degree of the burstiness of the traffic. We show the setting of the parameter b to zero result in a smooth traffic (poisson process) or setting to one result in batch arrivals process (when nodes respond to a query from the sink) . Of course, a very relevant parameter of our WSN model is the number of nodes, N, which merge together to feed the sink. fig 2 and fig 4 show that for ON/OFF processes with exponential ON and OFF times (T = l), the one node and two node are different in the intermediate region, but not significantly so. However, if the ON-times are TPT distributed (large T), even though the overall sink utilization, $\rho$, is kept at the same level, the mean packet delay in the one node and two node model both show sharp increases (blow-up points), this effect tells that it matters whether the incoming data streams from different nodes to a sink is modeled (and measured as one combined stream or as separate streams. As we see mPD and buffer overflow probability of the two node WSN model is smaller than the one node WSN model under the same utilization and arrival rate. In this paper, our simulations are capable of capturing the dynamic behavior of the WSN with pareto ON time distribution. This helps us gain useful insight of the network in the transient state, which may not be possibly obtained from theoretical analysis. Finally we find that by changing the WSN traffic model parameter can capture the most of WSN traffic application.

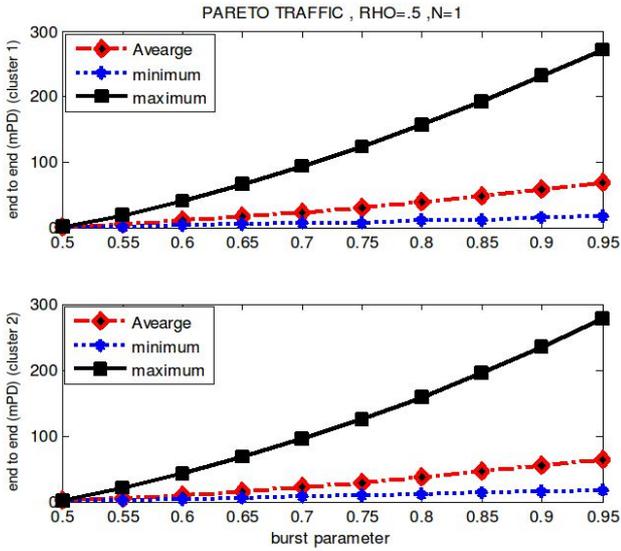

Figure 18: The mean packet delay with Pareto ON time distribution at N=1

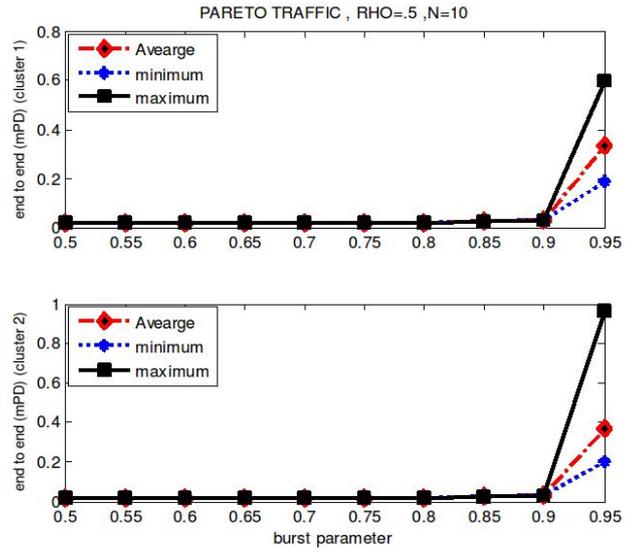

Figure 20: The mean packet delay with Pareto ON time distribution at N=10

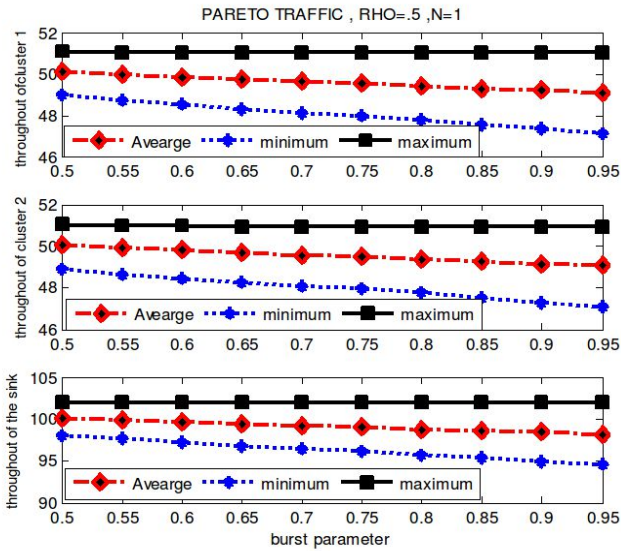

Figure 19: The throughout with Pareto ON time distribution at N=1

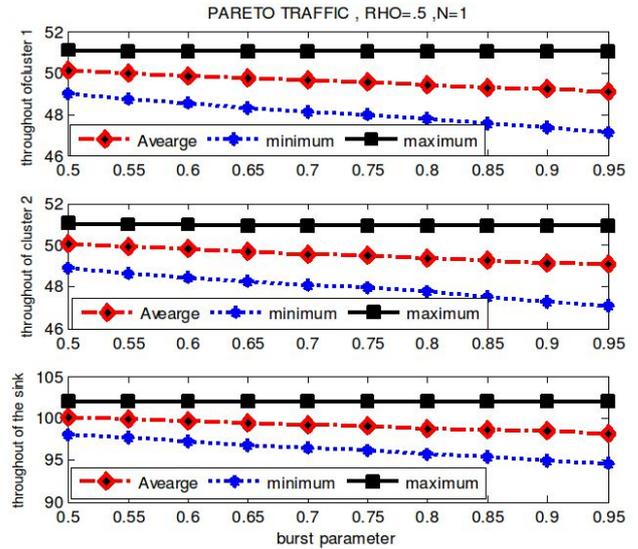

Figure 21: The throughout with Pareto ON time distribution at N=10

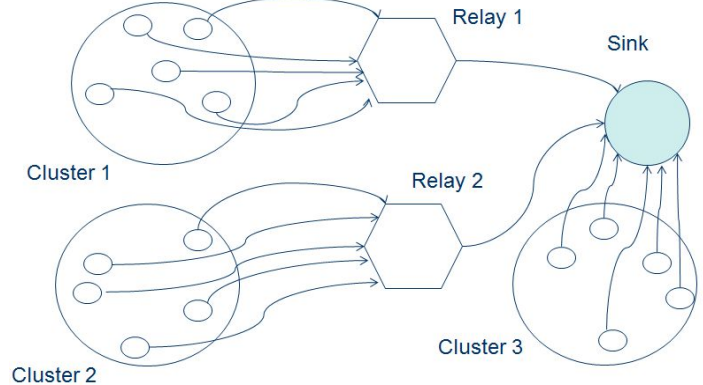

Figure 22: WSN with three clusters

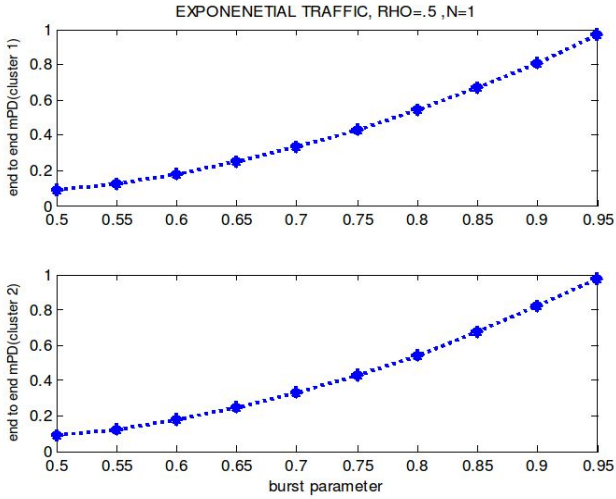

Figure 23: The mean packet delay with EXP ON time distribution at N=1

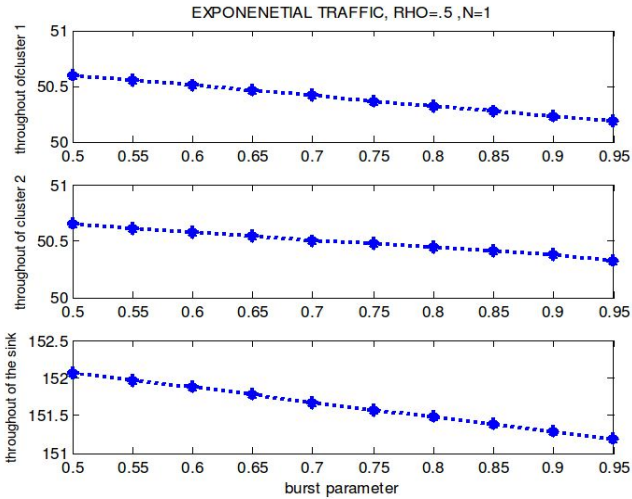

Figure 24: The throughout with EXP ON time distribution at N=1

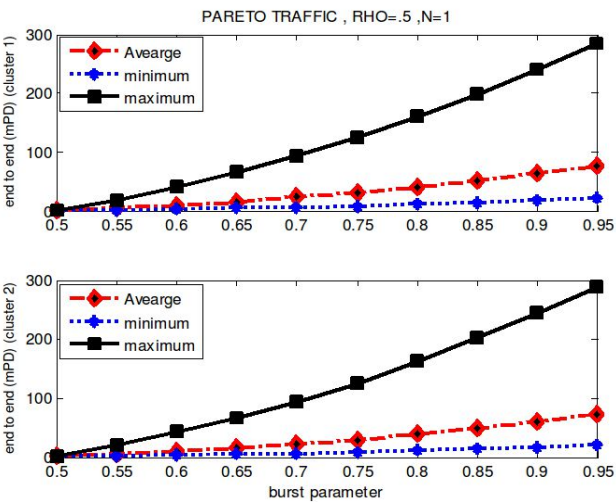

Figure 25: The mean packet delay with Pareto ON time distribution at N=1

Technology (RAIT), 2016 3rd International Conference on, IEEE, 2016, pp. 61–64.

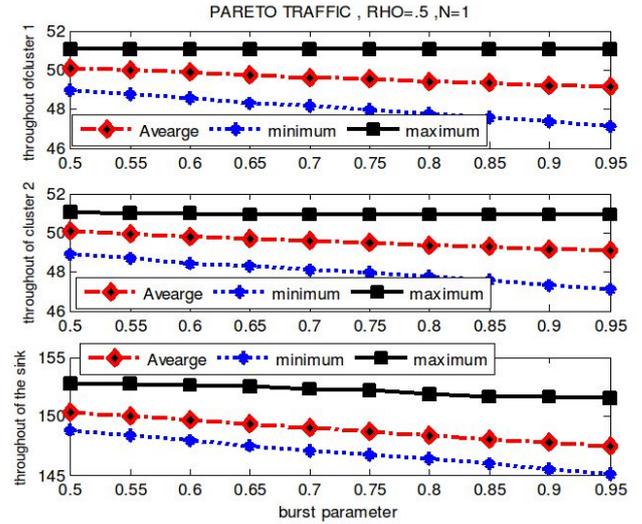

Figure 26: The throughout with Pareto ON time distribution at N=1